# The Bifurcation Growth Rate for the Robust Pattern Formation in the Reaction-Diffusion System on the Growing Domain

Shin Nishihara [1*] and Toru Ohira [1]

[1] Graduate School of Mathematics, Nagoya University.

Furocho, Chikusaku, Nagoya, 464-8602, Japan.

*Corresponding author: shin.kurokawa.c8@math.nagoya-u.ac.jp, ORCID: 0009-0003-2351-0644

**Abstract**

Among living organisms, there are species that change their patterns on their body surface during their growth process and those that maintain their patterns. Theoretically, it has been shown that large-scale species do not form distinct patterns. However, exceptionally, even large-scale species like giraffes form and maintain patterns, and previous studies have shown that the growth plays a crucial role in pattern formation and transition.

Here we show how the growth of the domain contributes to Turing bifurcation based on the reaction-diffusion system by applying the Gray-Scott model to the reaction terms, both analytically and numerically, focusing on the phenomenon of pattern formation and maintenance in large species like giraffes, where melanocytes are widely distributed. After analytically identifying the Turing bifurcation related to the growth rate, we numerically verify the pattern formation and maintenance in response to the finite-amplitude perturbations of the 'blue' state specific to the Gray-Scott model near the bifurcation. Furthermore, among pairs of the parameters that form Turing patterns in a reaction-diffusion system on a fixed domain, we determine a pair of the parameters that maximizes the growth rate for the Turing bifurcation in a reaction-diffusion system on a time-dependently growing domain. Specifically, we conduct a numerical analysis to pursue the pair of the parameters in the Turing space that can be the most robust in maintaining the patterns formed on the fixed domain, even as the domain grows. This study may contribute to specifically reaffirming the importance of growth rate in pattern formation and understanding patterns that are easy to maintain even during growth.

## 1. Introduction

During the growth process of living organisms, some species exhibit different patterns on their bodies as juveniles compared to adults, while other species maintain consistent patterns throughout their lives. Considering that many other elements change from juvenile to adult stages, it is possible that species with consistent patterns possess mechanisms for strictly controlling pattern formation. For example, in large animals like elephants, rhinoceroses, and hippopotamuses, which do not exhibit patterns, it is theoretically suggested and empirically observed that patterns are less likely to form on their bodies (Murray, 1988; Murray, 2013).

However, there are exceptional species, such as giraffes, which form distinctive patterns on their pale background color (Muller, 2017) from embryonic development through juvenile stages to adulthood. It has been reported that these patterns not only show significant variation among and within populations of the same subspecies but also vary by sex (Brand, 2007; Lee et al., 2018). Furthermore, subtle seasonal changes in the patterns have been observed; the patterns become lighter towards the end of the dry season and return to their original darker state in the rainy season (Brand, 2007). The patterns may also have ecological and physiological significance. The size of the patches is suggested to be related to reproductive success (Brand, 2007; Lee et al., 2018), and relatively large sweat glands exist beneath the patches (Mitchell-Frssaf and Skinner-Frssaf, 2004), with melanin potentially protecting these glands (Brenner and Hearing, 2008). Thus, the diverse pattern variations in giraffes could be a significant factor in their survival strategies.

Despite the extensive variation in giraffe's patterns, these patterns may be inherited from mother to offspring (Lee et al., 2018), suggesting the existence of some mechanism of pattern formation even in giraffes' rich variability. Remarkably, although pattern formation may have ecological and physiological significance, rare individuals without patterns have been reported. Albino individuals have been observed in the wild (Muller, 2017), and there have been reports of individuals with uniformly dark coats. For the latter, there is no similarity with the mother's pattern, and as far as we know, the reason for this uniform coat and lack of pattern formation remains unexplained.

In this study, we focus on the melanin pathway produced by melanocytes, which are more widely distributed in giraffes than in many other animals (Lee et al., 2018), and on the fact that individuals grow to a large size. We theoretically consider the Turing bifurcation in a time-dependent growing domain, distinguishing between cases where melanin patterns form and are maintained and where patterns do not form on a growing domain. Previous studies on giraffe pattern formation have used Reaction-Diffusion models and Clonal Mosaic models, and both models reproduce giraffe patterns despite differences in the required period for pattern formation (including pre-pattern formation) (Bard, 1981; Murray, 1981; Murray, 2013; Walter et al., 1998; Walter et al., 2001). In this paper, we adopt the former Reaction-Diffusion model for the reasons discussed later. Additionally, as melanin production decreases with growth (Sreedhar et al., 2020) and the target domain grows time-dependently, that indicates that we seek the bifurcation point based on a non-autonomous reaction-diffusion system. We use the Gray-Scott (GS) model (Gray and Scott, 1983; Gray and Scott, 1984; Gray, 1988; Pearson, 1993) for the reaction term for the following reasons: i) it can form a rich variety of patterns (Pearson, 1993), ii) the dilution term that appears in a growing domain's reaction-diffusion system (Nishihara and Ohira, 2024) does not significantly alter the original model's shape. While the maintenance of patterns on growing manifolds for Schnakenberg, Gierer-Meinhardt, and FitzHugh-Nagumo models (Schnakenberg, 1979; Gierer and Meinhardt, 1972; FitzHugh, 1961; Nagumo et al., 1962) have been theoretically and detailedly reported (Krause et al., 2019), in this study, we aim to theoretically and specifically investigate the bifurcation point of pattern maintenance and uniformity in a time-dependent growing domain using the GS model, considering the specific phenomenon of giraffe patterns and melanin production pathways.

## 2. Modeling and Analytical Solutions for the Bifurcation

In this paper, we analytically determine the bifurcation point at which patterns formed by the reaction-diffusion system on a growing domain are either maintained or disappear, and we use this Turing bifurcation point to numerically verify pattern transitions. Therefore, we first model and seek the analytical solution.

On the surface of a giraffe's body, melanin produced by widely distributed melanocytes forms patches that create distinctive patterns over a broad area, combined with a pale background color. For instance, if melanin is not produced, the giraffe is identified as albino (Muller, 2017). Conversely, if melanin is excessively produced uniformly over a wide area, individuals with uniformly dark fur can exist. Thus, this paper aims to construct a reaction-diffusion system focusing on melanin production.

Firstly, melanin is produced as tyrosine is converted according to a specific pathway (Cooksey et al., 1997; Eisenhofer et al., 2003; Fanet et al., 2021; Filimon and Negroiu, 2009; Riley, 1999; Sugumaran and Barek, 2016; Suzuki and Ichinose, 2005). Here, assuming that the GS model, which can form a variety of rich patterns (Pearson, 1993), contributes to the pathway of melanin production in certain species, we apply the GS model to the reaction term targeted in this study. Specifically, we assume that the areas with high values of $u$ (which eventually converts to melanin) form the giraffe's patches. The dimensionless reaction term is as follows:

$$F(u, v) \coloneqq -uv^2 + F(1 - u),$$
$$G(u, v) \coloneqq uv^2 - (F + K)v, \tag{1}$$

where $F$ and $K$ are positive constants. Additionally, melanin production does not continue throughout the lifespan but decreases with age (Sreedhar et al., 2020), so this needs to be reflected in the reaction term. Here, hypothetically, the decay depends on the area of the domain that expands over time, *i.e.*, it is inversely proportional. The reaction term is thus described as follows:

$$f(u,v) \coloneqq \mu(t)F(u,v) = \mu(t)\big(-uv^2 + F(1-u)\big)\,,$$
$$g(u,v) \coloneqq \mu(t)G(u,v) = \mu(t)\big(uv^2 - (F+K)v\big)\,,$$
$$\mu(t) \propto (Domain\ Area)^{-1}\,. \tag{2}$$

Furthermore, with the zero flux boundary conditions applied, the reaction-diffusion system on a growing square domain is described as follows (Nishihara and Ohira, 2024):

$$\frac{\partial u}{\partial t} = \frac{\phi_u}{r^2(t)}\left(\frac{\partial^2}{\partial x^2} + \frac{\partial^2}{\partial y^2}\right)u + f(u,v) - \frac{2r'(t)}{r(t)}u\,,$$

$$\frac{\partial v}{\partial t} = \frac{\phi_v}{r^2(t)}\left(\frac{\partial^2}{\partial x^2} + \frac{\partial^2}{\partial y^2}\right)v + g(u,v) - \frac{2r'(t)}{r(t)}v\,,$$

$$r'(t) \coloneqq \frac{\mathrm{d}r(t)}{\mathrm{d}t}\,, \text{and } \phi_u, \phi_v \text{ are constant diffusion coefficients}\,. \tag{3}$$

where, the area of the square domain at $t = 0$ is $L^2$, $x, y$ are time-independent spatial variables, and $r(t)$ is the time-dependent growth rate of the domain. Based on that, we define the growth rate $r(t)$, forming a pattern by the reaction-diffusion system on a fixed domain for $t < t_g$, and then examining whether the reaction-diffusion system maintains or loses the pattern on a growing domain for $t \geq t_g$ with the growth rate $r(t)$.

Specifically, we define as follows:

$$r(t) \coloneqq \begin{cases} 1, & t < t_g \\ \sqrt{\sigma t + 1}, & t \geq t_g \end{cases}\,,$$

$$\mu(t) \coloneqq \begin{cases} 1, & t < t_g \\ \dfrac{\kappa}{\big(Lr(t)\big)^2}, & t \geq t_g \end{cases}\,, \tag{4}$$

where $\sigma$ and $\kappa$ are positive constants.

From a theoretical perspective, the goal of this paper is to determine the Turing bifurcation point $\sigma_c$ concerning this growth rate $\sigma$ to predict pattern transitions. In the reaction-diffusion system on the fixed domain ($t < t_g$), appropriate $F, K, \phi_u,$ and $\phi_v$ that form Turing patterns based on the Turing mechanism are selected. Based on this parameter set, the system for $t \geq t_g$ is described as follows:

$$\frac{\partial u}{\partial t} = \frac{\phi_u}{r^2(t)}\left(\frac{\partial^2}{\partial x^2} + \frac{\partial^2}{\partial y^2}\right)u + \mu(t)\big(-uv^2 + F(1-u)\big) - \frac{2r'}{r}u\,,$$

$$\frac{\partial v}{\partial t} = \frac{\phi_v}{r^2(t)}\left(\frac{\partial^2}{\partial x^2} + \frac{\partial^2}{\partial y^2}\right)v + \mu(t)\big(uv^2 - (F+K)v\big) - \frac{2r'}{r}v\,. \tag{5}$$

For simplicity in the discussion, let $\kappa = L^2$, and additionally,

$$\tau \coloneqq \frac{1}{\sigma}\log(\sigma t + 1)\,, \tag{6}$$

by which the non-autonomous reaction-diffusion system can be converted into an autonomous reaction-diffusion system (see Appendix-A for details):

$$\frac{\partial u}{\partial \tau} = \phi_u \left( \frac{\partial^2}{\partial x^2} + \frac{\partial^2}{\partial y^2} \right) u + \hat{f}(u,v) \, , \text{where } \hat{f}(u,v) \coloneqq -uv^2 + F - (F+\sigma)u \, ,$$

$$\frac{\partial v}{\partial \tau} = \phi_v \left( \frac{\partial^2}{\partial x^2} + \frac{\partial^2}{\partial y^2} \right) v + \hat{g}(u,v) \, , \text{where } \hat{g}(u,v) \coloneqq uv^2 - (F+\sigma+K)v \, . \tag{7}$$

Thus, for the Turing pattern formed on the fixed domain to be maintained according to the Turing mechanism even during domain growth, the following conditions must be satisfied:

(A) The condition for the existence of the 'blue' state (Mazin et al., 1996) $(\hat{u}^*, \hat{v}^*)$ :

$$D(\sigma) \coloneqq F^2 - 4(F+\sigma)(F+\sigma+K)^2 > 0 \, . \tag{8}$$

(B) The condition for the stability of the equilibrium state $(\hat{u}^*, \hat{v}^*)$ :

$$\text{tr}\,\hat{S}^* < 0, \text{and} \det \hat{S}^* > 0 \, , \text{where}$$

$$\hat{S}^* \coloneqq \begin{pmatrix} \hat{f}_u^* & \hat{f}_v^* \\ \hat{g}_u^* & \hat{g}_v^* \end{pmatrix} = \begin{pmatrix} -\hat{v}^{*2} - (F+\sigma) & -2(F+\sigma+K) \\ \hat{v}^{*2} & (F+\sigma+K) \end{pmatrix} \, . \tag{9}$$

(C) The Turing Instability conditions:

$$TIC_1(\sigma) \coloneqq \phi_v \hat{f}_u^* + \phi_u \hat{g}_v^* > 0 \, ,$$
$$TIC_2(\sigma) \coloneqq TIC_1^2(\sigma) - 4\phi_u \phi_v (\det \hat{S}^*) > 0 \, . \tag{10}$$

By organizing these conditions, the bifurcation point concerning the growth rate $\sigma_c$ can be obtained (see Appendix-B for details):

$$\sigma_c = \begin{cases} \sqrt[4]{F^2 K} - (F+K), & (K-F) \geq \sigma_D \\ \sigma_D, & (K-F) < \sigma_D \end{cases} \tag{11}$$

where $\sigma_D$ is the bifurcation point concerning the positive upper limit value of $\sigma$ obtained from the cubic inequality of condition (A), and note that we later define $\sqrt[4]{F^2 K} - (F+K)$ and $(K-F)$, as $\sigma_T$ and $\sigma_{BT}$, respectively (see Appendix-B for details). In other words, if $\sigma_{BT} - \sigma_D$ is non-negative then the Turing bifurcation on a growing domain is the Hopf bifurcation, while if $\sigma_{BT} - \sigma_D$ is negative then the Turing bifurcation is the saddle-node bifurcation. Therefore, it is predicted by linear theory that the pattern formed will be maintained within this growth rate bifurcation point range. Conversely, if this growth rate bifurcation point is exceeded, it suggests the possibility of the Turing pattern formed disappearing.

On the other hand, concerning $G(u,v)$ adopted above, in reality, a basal feed rate $(\varepsilon)$, albeit slight, may exist in melanin production as follows:

$$G'(u,v) \coloneqq G(u,v) + \varepsilon \, . \tag{12}$$

Here, for the purpose of analytically determining the bifurcation point, $\varepsilon = 0$ is assumed, but this $G'(u,v)$ will be numerically verified with a small $\varepsilon$.

## 3. Result

To verify that the analytically determined growth rate bifurcation point can reproduce pattern maintenance and disappearance through numerical analysis, we use the FTCS (Forward Time Centered Space) scheme for the system (5). Previous studies on giraffe pattern formation based on reaction-diffusion systems have set a threshold for melanin production (Bard, 1981; Murray, 1988). However, intuitively, setting this threshold extremely low can reproduce uniformly dark fur, so in this study, we aim to reproduce giraffe pattern formation without setting this threshold. It should be noted that understanding melanin production starts once a certain threshold is exceeded is common (Murray, 2013), and the absence of a threshold in this study is merely for simplification of the discussion. Furthermore, in the parameter setting of the GS model, we ensure $\phi_u/\phi_v \gg 1$ to avoid spatio-temporal chaos (Nishiura and Ueyama, 2001) and also avoid the vicinity of the Bogdanov-Takens point, given by $F = 4\sqrt{K^3}$ (see also Appendix-B for more information), to focus the discussion.

Based on these considerations, for instance, we set the parameters that can form patterns according to the Turing mechanism as follows:

$$\phi_u = 0.2, \phi_v = 0.1, F = 0.089, \text{ and } K = 0.06. \tag{13}$$

First, we confirm that patterns form on a fixed square domain ($L = 100$). Specifically, we verify pattern formation at $t < t_g = 1,000$.

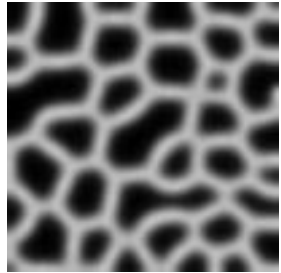

**Figure 1**: Numerical analysis results for $u$ using the FTCS scheme based on the system (5) and the parameter sets (4), (13). The side length of the square domain is $L = 100$, and numerical analysis is performed until $t = t_g = 1,000$. In the figure, dark color (black) represents high values of $u$, and light color (white) represents low values (see also Appendix-D).

Next, we confirm the patterns formed based on the growing domain system (5), and the above parameter set (13) satisfies the following condition:

$$(K - F) < \sigma_D \tag{14}$$

Therefore, the growth rate bifurcation point is as follows (see Appendix-B for details):

$$\sigma_c = \sigma_D \approx 8.9 \times 10^{-5} \tag{15}$$

Based on the definition of $r(t)$ representing the growth rate of the domain, we run the system (5) until $t = 33,000 > t_g$ so that the side length of the domain approximately doubles (area increases about fourfold).

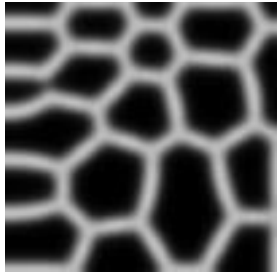

**Figure 2**: Numerical analysis results for $u$ using the FTCS scheme based on the system (5) and the parameter sets (4), (13), (15). At $t = t_g = 1,000$, the side length of the square domain is $L = 100$, and numerical analysis is performed until $t = 33,000$, so the side length of the square domain reaches approximately twice ($L = 200$). In the figure, dark color (black) represents high values of $u$, and light color (white) represents low values (see also Appendix-D).

As predicted by linear theory, it is confirmed that the formed pattern is maintained even if the target domain doubles, *i.e.*, the area increases about fourfold, at the growth rate bifurcation point. Furthermore, as mentioned earlier, realistically, it is possible that replacing $G(u,v)$ with $G'(u,v)$ might more accurately represent the phenomenon in melanin production. Therefore, we replace $G(u,v)$ with $G'(u,v)$, set $\varepsilon = 0.0001$, and run the system (5).

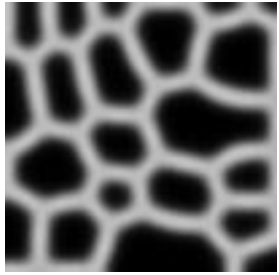

**Figure 3**: Numerical analysis results for $u$ using the FTCS scheme based on the system (5) and the parameter sets (4), (13), (15). At $t = t_g = 1,000$, the side length of the square domain is $L = 100$, and numerical analysis is performed until $t = 33,000$, so the side length of the square domain reaches approximately twice ($L = 200$). Note that, in the system (5), $G(u,v)$ is replaced with $G'(u,v)$, and $\varepsilon = 0.0001$. In the figure, dark color (black) represents high values of $u$, and light color (white) represents low values (see also Appendix-D).

As shown above, the numerical analysis result incorporating a basal feed rate with a small $\varepsilon$ yields similar result to when $\varepsilon = 0$. Larger values of $\varepsilon$ will be discussed later.

On the other hand, when numerical analysis is performed using $\sigma$ that exceeds the growth rate bifurcation point $\sigma_c$ until the side length of the square domain reaches approximately twice ($L = 200$), it is confirmed that the formed pattern disappears as follows.

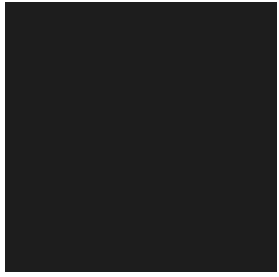

**Figure 4**: Numerical analysis results for $u$ using the FTCS scheme based on the system (5) and the parameter sets (4), (13). At $t = t_g = 1{,}000$, the side length of the square domain is $L = 100$. Setting $\sigma = 0.01$, which greatly exceeds the growth rate bifurcation point $\sigma_c$, numerical analysis is performed until $t = 1{,}300$, and the side length of the square domain reaches approximately twice ($L = 200$). In the figure, dark color (black) represents high values of $u$, and light color (white) represents low values (see also Appendix-D).

## 4. Conclusion

This study aims to analytically determine the Turing bifurcation point for the maintenance or disappearance of pattern formation based on the Turing mechanism in the non-autonomous reaction-diffusion system on a growing domain. To achieve this, we specifically assume the growth rate $r(t)$ of the domain and the time-decaying reaction term $\mu(t)$, confirm the existence of these bifurcation points, and analytically identify them. Additionally, while we use numerical analysis to demonstrate that the growth rate bifurcation point theoretically derived can maintain the pattern, the numerical analysis confirms that patterns disappear when the growth rate exceeds this bifurcation point. In terms of pattern of animals' body surface, this study suggests that even large animals like giraffes can form and maintain patterns on their bodies during growth. Furthermore, it theoretically demonstrates that by changing a single parameter, the growth rate, it is possible to control the maintenance and disappearance of these patterns. Rare occurrences of spotless giraffes could theoretically be explained by growth rates significantly exceeding the analytically determined growth rate bifurcation point. Thus, while previous studies have shown that the scale of the domain significantly impacts pattern formation (Murray, 1988; Murray, 2013), this study suggests that the growth rate of the domain scale also plays a crucial role in pattern formation. In fact, despite the significant difference in scales between giraffes and lions, it has been reported that the former maintains its pattern during growth while the latter loses it (Walter et al., 2001), implying that not only the fixed scale but also the growth rate contributes.

In this study, numerical analysis is performed using the aforementioned parameter set. Since the growth rate bifurcation point $\sigma_c$ can be expressed solely in terms of $F$ and $K$ (note that $\sigma_D$ is also expressed only in terms of $F$ and $K$), it is theoretically predicted that a growth rate bifurcation point $\sigma_c$ exists by fixing $\phi_u$ and $\phi_v$ and selecting appropriate $F$ and $K$. Moreover, it is predicted that there exists a combination of $F$ and $K$ that maximizes the growth rate bifurcation point $\sigma_c$, making it the most robust for pattern maintenance during the domain growth. As shown in the following numerical analysis results (this numerical analysis is performed with $10^{-4}$ as the smallest unit for $F$ and $K$), there exists a combination of $F$ and $K$ that maximizes the growth rate bifurcation point $\sigma_c$ (marked by the green star in the figure), potentially suggesting this type of patterns may be easy to be maintained even during growth after being formed during earlier stages. Considering the value of $\sigma_c$ given in the equation (15) (see also Appendix-C for more information), the pattern formed by this pair of $(F, K)$ identifying $\sigma_c$ (similar to the pattern on the body surface of giraffes) is likely to be relatively well maintained. The theoretical robustness of the growth rate bifurcation point $\sigma_c$ in maintaining patterns, combined with the pattern formation and maintenance in response to finite-amplitude perturbations of the blue state characteristic of the GS model, theoretically demonstrates that a reaction-diffusion system applying the time-dependent GS model on a growing domain provides robustness to already formed patterns, even during the growth.

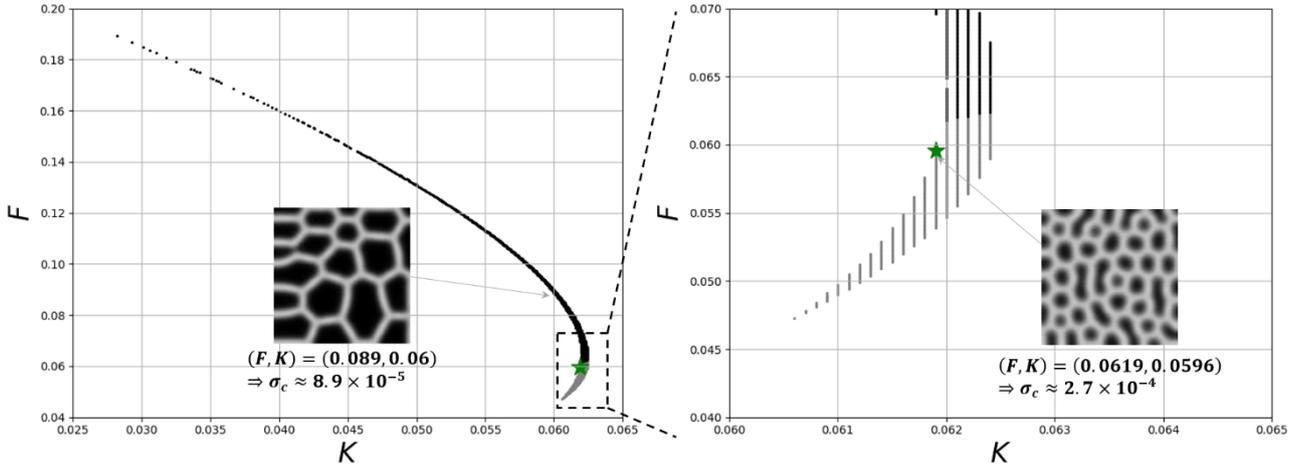

**Figure 5**: This figure plots the combinations of $F$ and $K$ that satisfy the Turing instability conditions with fixed $\phi_u = 0.2$ and $\phi_v = 0.1$ on a fixed domain. Black dots represent combinations of $F$ and $K$ where $(K - F) < \sigma_D$, meaning $\sigma_c$ equals $\sigma_D$. Grey dots represent the combinations where $(K - F) \geq \sigma_D$, meaning $\sigma_c$ equals $\sigma_T$. The green star indicates the combination $(F_{robust}, K_{robust}) = (0.0619, 0.0596)$ that maximizes $\sigma_c$ ($\sigma_{robust} = \sigma_T \approx 0.00027$), making it the most robust for pattern maintenance against growth rate fluctuations. The pattern diagrams corresponding to the parameter sets (13), (15), and $(F_{robust}, K_{robust}, \sigma_{robust})$ are attached. At $t = t_g = 1,000$, the side length of the square domain is $L = 100$, and for the latter pattern diagram, numerical analysis is performed until $t = 12,000$ using $(F_{robust}, K_{robust}, \sigma_{robust})$, resulting in the side length of the square domain reaching approximately twice, $L = 200$. The right side of the figure is an enlarged view of the region outlined by the dotted lines on the left side (see Appendix-C for more details). (Star in Color)

## 5. Discussion

In this paper, it is important to note that the mechanisms of pattern formation on the giraffe's skin are not demonstrated directly, as several assumptions are set, such as the growth rate of the domain $r(t)$, the decay of the reaction $\mu(t)$, and the GS model adopted for the pathway involved in melanin production. For example, while l-Dopa functions as a coenzyme to enhance the catalytic efficiency and promote reactions in the melanin production pathway (Eisenhofer et al., 2003), it is somewhat optimistic to consider that the GS model itself completely matches the following pathway: (1) Tyrosinase promotes the conversion of Tyrosine to Dopaquinone, (2) Dopaquinone is converted to Cyclodopa, (3) Dopaquinone and Cyclodopa are nonenzymatically converted to l-Dopa and Dopachrome, (4) Dopachrome is eventually metabolized into melanin, (5) l-Dopa is converted to Dopaquinone by Tyrosinase, and (6) l-Dopa is also converted from Tyrosine by Tyrosine hydroxylase and the coenzyme Tetrahydrobiopterin (Cooksey et al., 1997; Eisenhofer et al., 2003; Fanet et al., 2021; Filimon and Negroiu, 2009; Riley, 1999; Sugumaran and Barek, 2016; Suzuki and Ichinose, 2005). Nevertheless, discussing theoretical predictions obtained through assumptions is also important. Particularly, for organisms like giraffes, which have i) axons of specific lengths (Howe, 2005) and ii) the need to circulate blood throughout their long bodies, neurotransmitters and hormones such as Dopamine and Epinephrine, which are converted from l-Dopa (Eisenhofer, 2003), are considered crucial. If l-Dopa is not properly converted to these substances and the amount of l-Dopa increases, it means that $\varepsilon$ in $G'(u,v)$ would be large when $v$ is regarded as l-Dopa. As shown below, numerical analysis results confirm a phenomenon where $u$ (melanin) is uniformly distributed:

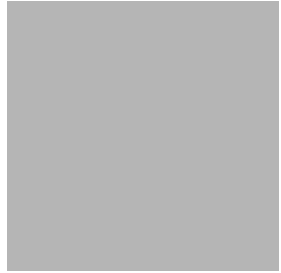

**Figure 6**: Results for $u$ obtained through numerical analysis using the FTCS scheme based on the system (5) and the parameter sets (4), (13), (15). The side length of the square domain is $L = 100$, and numerical analysis is performed up to $t = t_g = 1{,}000$. Note that $G'(u,v)$ is used instead of $G(u,v)$ in the system (5), with $\varepsilon = 0.001$. In the figure, dark color (black) indicates high values of $u$, and light color (white) indicates low values, showing a uniform intermediate shade in this case (see also Appendix-D).

The temporal decrease in Epinephrine secretion has been reported (Seals and Esler, 2000; Stamou et al., 2022), reflecting the time decay in the reaction term for $v$ (*e.g.*, l-Dopa) in the system (5). Although these are theoretical considerations based on assumptions, it is possible to imagine that growth, melanin, neurotransmitters, and hormones interact complexly and indirectly, which is a topic for future research.

In this study, we theoretically derived the growth rate bifurcation point $\sigma_c$. However, numerical analysis shows that even with a growth rate $\sigma$ slightly exceeding $\sigma_c$, the formed pattern is maintained:

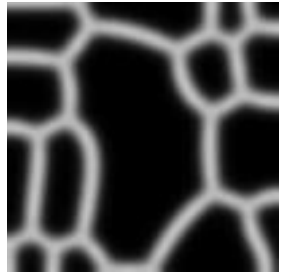

**Figure 7**: Results for $u$ obtained through numerical analysis using the FTCS scheme based on the system (5) and the parameter sets (4), (13). At $t = t_g = 1,000$, the side length of the square domain is $L = 100$, and numerical analysis is performed up to $t = 31,000$ using a growth rate $\sigma = 0.0001$, slightly exceeding $\sigma_c$, resulting in the side length of the square domain about twice, $L = 200$. In the figure, dark color (black) indicates high values of $u$, and light color (white) indicates low values (see also Appendix-D).

The following reasons are speculated for this discrepancy: i) the theoretically derived growth rate bifurcation point $\sigma_c$ is based on a linear approximation of linear theory, causing differences, ii) the GS model forms patterns from finite-amplitude perturbations (Pearson, 1993), so in the systems (5) and (7) at $t \geq t_g$, the previously formed Turing patterns correspond to finite-amplitude perturbations, forming and maintaining patterns in response to the blue state perturbations of finite amplitude, separate from the Turing mechanism (Mazin et al., 1996), and iii) in the vicinity of the growth rate bifurcation point $\sigma_c$, the reaction and dilution terms cause the solution to stay in the vicinity of the spatially non-uniform stationary solution, *i.e.*, the patterns remain as '*lingering*' patterns (Mimura, 2006) (see Appendix-E for more information). As detailed in Appendix-B, considering that the conditions required for pattern maintenance are monotonically increasing or decreasing, increasing the value of $\sigma$ beyond the bifurcation point $\sigma_c$ results in the disappearance of patterns, as shown below. When $\sigma$ slightly exceeds the bifurcation point (case (**b**) below), the system approximately retains the shape of the GS model, so patterns may be maintained by the reason: ii). However, when $\sigma$ increases to the extent that the system no longer resembles the GS model (cases (**c**) and (**d**) below), even this reason ceases to apply.

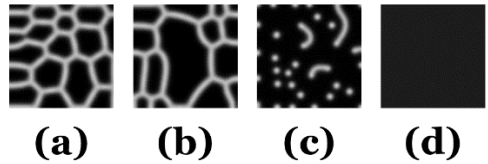

(a)      (b)      (c)      (d)

**Figure 8**: Results for $u$ obtained through numerical analysis using the FTCS scheme based on the system (5) and the parameter sets (4), (13). Numerical analysis is performed up to a time $t$ where the side length of the square domain at $t = t_g = 1,000$ (with $L = 100$) approximately doubles for each value of $\sigma$: (**a**) $(\sigma, t) = (\sigma_c \approx 0.000089, 33,000)$, (**b**) $(\sigma, t) = (0.0001, 31,000)$, (**c**) $(\sigma, t) = (0.001, 4,000)$, (**d**) $(\sigma, t) = (0.01, 1,300)$. In the figure, dark color (black) indicates high values of $u$, and light color (white) indicates low values.

## 6. Author contributions

**Shin Nishihara:** Conceptualization, Methodology, Software, Validation, Formal analysis, Investigation, Resources, Data Curation, Writing - Original Draft, Writing - Review and Editing, and Visualization

**Toru Ohira:** Resources, Writing - Review and Editing, Supervision, Project administration, and Funding acquisition

## 7. Acknowledgements and Funding

This work was supported by JSPS Topic-Setting Program to Advance Cutting-Edge Humanities and Social Sciences Research Grant Number JPJS00122674991, and by Ohagi Hospital, Hashimoto, Wakayama, Japan.

## 8. References

J. B. L. Bard. A model for generating aspects of zebra and other mammalian coat patterns. *Journal of Theoretical Biology*, 93(2):363-385, 1981. doi: 10.1016/0022-5193(81)90109-0

R. Brand. Evolutionary Ecology of Giraffes (Giraffa camelopardalis) in Etosha National Park, Namibia. *PhD Thesis, Newcastle University*, 2007. doi: http://theses.ncl.ac.uk/jspui/handle/10443/1461

M. Brenner and V. J. Hearing. The Protective Role of Melanin Against UV Damage in Human Skin. *Photochemistry and Photobiology*, 84(3):539-549, 2008. doi: 10.1111/j.1751-1097.2007.00226.x

C. J. Cooksey, P. J. Garratt, E. J. Land, S. Pavel, C. A. Ramsden, P. A. Riley and N. P. M. Smit. Evidence of the Indirect Formation of the Catecholic Intermediate Substrate Responsible for the Autoactivation Kinetics of Tyrosinase. *Journal of Biological Chemistry*, 272(42):26226-26235, 1997. doi: 10.1074/jbc.272.42.26226

G. Eisenhofer, H. Tan, C. Holmes, J. Matsunaga, S. Roffler-Tarlov and V. J. Hearing. Tyrosinase: a developmentally specific major determinant of peripheral dopamine. *The FASEB Journal*, 17(10):1195-1376, 2003. doi: 10.1096/fj.02-0736com

H. Fanet, L. Capuron, N. Castanon, F. Calon and S. Vancassel. Tetrahydrobioterin (BH4) Pathway: From Metabolism to Neuropsychiatry. *Current Neuropharmacology*, 19(5):591-609, 2021. doi:

10.2174/1570159X18666200729103529

A. Filimon and G. Negroiu. DOPACHROME TAUTOMERASE – AN OLD PROTEIN WITH NEW FUNCTIONS. *Rom J Biochem*, 299:36-52, 2009.

R. FitzHugh. Impulses and Physiological States in Theoretical Models of Nerve Membrane. *Biophysical Journal*, 1(6):445-466, 1961. doi: 10.1016/S0006-3495(61)86902-6

A. Gierer and H. Meinhardt. A theory of biological pattern formation. *Kybernetik*, 12:30-39, 1972. doi: 10.1007/BF00289234

P. Gray. Instabilities and oscillations in chemical reactions in closed and open systems. *Proceedings of the Royal Society A*, 415(1848), 1988. doi: 10.1098/rspa.1988.0001

P. Gray and S. K. Scott. Autocatalytic reactions in the isothermal, continuous stirred tank reactor: Isolas and other forms of multistability. *Chemical Engineering Science*, 38(1):29-43, 1983. doi: 10.1016/0009-2509(83)80132-8

P. Gray and S. K. Scott. Autocatalytic reactions in the isothermal, continuous stirred tank reactor: Oscillations and instabilities in the system $A + 2B \rightarrow 3B$; $B \rightarrow C$. *Chemical Engineering Science*, 39(6):1087-1097, 1984. doi: 10.1016/0009-2509(84)87017-7

C. L. Howe. Modeling the signaling endosome hypothesis: Why a drive to the nucleus is better than a (random) walk. *Theoretical Biology and Medical Modelling*, 2(43), 2005. doi: 10.1186/1742-4682-2-43

A. L. Krause, M. A. Ellis and R. A. Van Gorder. Influence of Curvature, Growth, and Anisotropy on the Evolution of Turing Patterns on Growing Manifolds. *Bulletin of Mathematical Biology*, 81:759-799, 2019. doi: 10.1007/s11538-018-0535-y

D. E. Lee, D. R. Cavener and M. L. Bond. Seeing spots: quantifying mother-offspring similarity and assessing fitness consequences of coat pattern traits in a wild population of giraffes (Giraffa camelopardalis). *PeerJ*, 6:e5690, 2018. doi: 10.7717/peerj.5690

W. Mazin, K. E. Rasmussen, E. Mosekilde, P. Borckmans and G. Dewel. Pattern formation in the bistable Gray-Scott model. *Mathematics and Computers in Simulation*, 40(3-4):371-396, 1996. doi: 10.1016/0378-4754(95)00044-5

M. Mimura. Mathematics of Nonlinear-Nonequilibrium Phenomena 4: Pattern Formation and its Dynamics. *University of Tokyo Press*, 2006.

G. Mitchell-Frssaf and J. D. Skinner-Frssaf. Giraffe Thermoregulation: a review. *Transactions of the Royal Society of South Africa*, 59(2):109-118, 2004. doi: 10.1080/00359190409519170

Z. Muller. White giraffes: The first record of vitiligo in a wild adult giraffe. *African Journal of Ecology*, 55(1):118-123, 2017. doi: 10.1111/aje.12323

J. D. Murray. A Pre-pattern formation mechanism for animal coat markings. *Journal of Theoretical Biology*, 88(1):161-199, 1981. doi: 10.1016/0022-5193(81)90334-9

J. D. Murray. How the Leopard Gets Its Spots. *Scientific American*, 258(3):80-87, 1988. doi: 10.1038/scientificamerican0388-80

J. D. Murray. Mathematical Biology II: Spatial Models and Biomedical Applications. *Springer*, 2013.

J. Nagumo, S. Arimoto and S. Yoshizawa. An Active Pulse Transmission Line Simulating Nerve Axon. *Proceedings of the IRE*, 50(10):2061-2070, 1962. doi: 10.1109/JRPROC.1962.288235

S. Nishihara and T. Ohira. The mechanism of pattern transitions between formation and dispersion. *Journal of Theoretical Biology*, 581, 2024. doi: 10.1016/j.jtbi.2024.111736

Y. Nishiura and D. Ueyama. Spatio-temporal chaos for the Gray–Scott model. Physica D: *Nonlinear Phenomena*, 150(3-4):137-162, 2001. doi: 10.1016/S0167-2789(00)00214-1

J. E. Pearson. Complex Patterns in a Simple System. *Science*, 261(5118):189-192, 1993. doi: 10.1126/science.261.5118.189

P. A. Riley. The great DOPA mystery: the source and significance of DOPA in phase I melanogenesis. *Cellular and Molecular Biology*, 45(7):951-960, 1999.

J. Schnakenberg. Simple chemical reaction systems with limit cycle behaviour. *Journal of Theoretical Biology*, 81:389-400, 1979. doi: 10.1016/0022-5193(79)90042-0

D. R. Seals and M. D. Esler. Human ageing and the sympathoadrenal system. *Journal of Physiology*, 528(3):407-417, 2000. doi: 10.1111/j.1469-7793.2000.00407.x

A. Sreedhar, L. Aguilera-Aguirre and K. K. Singh. Mitochondria in skin health, aging, and disease. *Cell Death & Disease*, 11(444), 2020. doi: 10.1038/s41419-020-2649-z

M. I. Stamou, C. Colling and L. E. Dichtel. Adrenal aging and its effects on the stress response and immunosenescence. *Maturitas*, 168:13-19, 2022. doi: 10.1016/j.maturitas.2022.10.006

M. Sugumaran and H. Barek. Critical Analysis of the Melanogenic Pathway in Insects and Higher Animals. *International Journal of Molecular Sciences*, 17(10):1753, 2016. doi: 10.3390/ijms17101753

T. Suzuki and H. Ichinose. Recent Progress in the Study of Tetrahydrobiopterin. *Vitamins*, 79(2):87-95, 2005. doi: 10.20632/vso.79.2_87

M. Walter, A. Fournier and M. Reimers. Clonal Mosaic Model for the Synthesis of Mammalian Coat Patterns. *Graphics Interface*, 82-91, 1998. doi: 10.20380/GI1998.11

M. Walter, A. Fournier and D. Menevaux. Integrating shape and pattern in mammalian models. *Computer Graphics (SIGGRAPH '01: Proceedings of the 28th annual conference on Computer graphics and interactive techniques)*, 317-326, 2001. doi: 10.1145/383259.383294

## Appendix

### A. Conversion to the Autonomous Reaction-Diffusion System

We convert the non-autonomous reaction-diffusion system (5) into the autonomous reaction-diffusion system. First, we describe each term on the RHS of the system (5). The diffusion term is described as follows based on the definition of $r(t)$:

$$\frac{\phi_u}{\sigma t + 1} \left( \frac{\partial^2}{\partial x^2} + \frac{\partial^2}{\partial y^2} \right) u,$$

$$\frac{\phi_v}{\sigma t + 1} \left( \frac{\partial^2}{\partial x^2} + \frac{\partial^2}{\partial y^2} \right) v. \tag{A1}$$

The reaction terms are described as follows based on the definitions of $\kappa$ and $r(t)$:

$$\frac{1}{\sigma t + 1} F(u, v),$$

$$\frac{1}{\sigma t + 1} G(u, v). \tag{A2}$$

The dilution terms are described as follows based on the definition of $r(t)$:

$$\frac{2r'}{r} = 2 \cdot \frac{1}{2} \cdot \frac{\sigma}{\sqrt{\sigma t + 1} \cdot \sqrt{\sigma t + 1}} = \frac{\sigma}{\sigma t + 1}$$

$$\Rightarrow -\frac{\sigma}{\sigma t + 1} u, \text{ and } -\frac{\sigma}{\sigma t + 1} v. \tag{A3}$$

Next, regarding the LHS of the system (5), based on the definition of $\tau$, it is as follows:

$$\frac{d\tau}{dt} = \frac{1}{\sigma} \cdot \frac{\sigma}{\sigma t + 1} = \frac{1}{\sigma t + 1}. \tag{A4}$$

Therefore, when we organize the above, the system (5) is converted as follows:

$$\frac{\partial u}{\partial \tau} = \phi_u \left( \frac{\partial^2}{\partial x^2} + \frac{\partial^2}{\partial y^2} \right) u + F(u, v) - \sigma u,$$

$$\frac{\partial v}{\partial \tau} = \phi_v \left( \frac{\partial^2}{\partial x^2} + \frac{\partial^2}{\partial y^2} \right) v + G(u, v) - \sigma v. \tag{A5}$$

### B. Derivation of the Growth Rate Bifurcation Point

The linearly stable equilibrium state in the system (7) is described as follows:

$$(\hat{u}^*, \hat{v}^*) = \left( \frac{F - \sqrt{F^2 - 4(F + \sigma)(F + \sigma + K)^2}}{2(F + \sigma)}, \frac{F + \sqrt{F^2 - 4(F + \sigma)(F + \sigma + K)^2}}{2(F + \sigma + K)} \right) \tag{B1}$$

For this linearly stable equilibrium state to exist in the system (7), the following condition must be satisfied:

$$D(\sigma) := F^2 - 4(F + \sigma)(F + \sigma + K)^2 > 0 . \tag{B2}$$

Since $D(\sigma)$ is monotonically decreasing and, under the condition where the region is not growing, $D := F^2 - 4F(F + K)^2 > 0$, there exists a $\sigma_D > 0$ where $D(\sigma_D) = 0$.

Next, the Jacobian required to find the conditions for this equilibrium state to be linearly stable is described as follows:

$$\hat{S}^* := \begin{pmatrix} \hat{f}_u^* & \hat{f}_v^* \\ \hat{g}_u^* & \hat{g}_v^* \end{pmatrix} = \begin{pmatrix} -\hat{v}^{*2} - (F + \sigma) & -2(F + \sigma + K) \\ \hat{v}^{*2} & (F + \sigma + K) \end{pmatrix} \tag{B3}$$

Here, $\hat{f}_u^*$ represents $\partial \hat{f}/\partial u(\hat{u}^*, \hat{v}^*)$, and similarly for the other terms. Since $\hat{v}^*$ decreases monotonically with increasing $\sigma$,

$$\mathrm{tr}\, \hat{S}^* = -\hat{v}^{*2} + K \tag{B4}$$

is monotonically increasing with increasing $\sigma$. Therefore, there must be a $\sigma_T$ where $\mathrm{tr}\, \hat{S}^* = 0$. If $2(F + \sigma_D + K)^3 \geq F^2$, i.e.,

$$(K - F) \geq \sigma_D \tag{B5}$$

then, solving $\mathrm{tr}\, \hat{S}^* = 0$ yields,

$$\sigma_T = \sqrt[4]{F^2 K} - (F + K) . \tag{B6}$$

On the other hand, if $2(F + \sigma_D + K)^3 < F^2$, i.e.,

$$(K - F) < \sigma_D \tag{B7}$$

then, there is no case where $\mathrm{tr}\, \hat{S}^* \geq 0$, so as long as a linearly stable equilibrium state exists, $\mathrm{tr}\, \hat{S}^* < 0$ must be satisfied. Furthermore,

$$\det \hat{S}^* = F\hat{v}^* - 2(F + \sigma)(F + \sigma + K) \tag{B8}$$

is also monotonically decreasing with increasing $\sigma$, and under the condition where the region is not growing, the determinant of the Jacobian is positive. In addition:

$$\sigma = \sigma_D \Rightarrow \det \hat{S}^* = \frac{F^2 - 4(F + \sigma)(F + \sigma + K)^2}{2(F + \sigma + K)} = 0 . \tag{B9}$$

Therefore, as long as a linearly stable equilibrium state exists, $\det \hat{S}^* > 0$.

Finally, regarding the following condition of the Turing mechanism:

$$TIC_1(\sigma) := \phi_v \hat{f}_u^* + \phi_u \hat{g}_v^* = -\frac{\phi_v F}{F + \sigma + K} \hat{v}^* + \phi_u(F + \sigma + K) > 0 \tag{B10}$$

$TIC_1(\sigma)$ is monotonically increasing with increasing $\sigma$, and since the Turing instability conditions are satisfied under the condition where the region is not growing, this condition is satisfied if $\sigma > 0$. Regarding another condition of the Turing mechanism:

$$TIC_2(\sigma) := TIC_1^2(\sigma) - 4\phi_u \phi_v (\det \hat{S}^*) > 0 \tag{B11}$$

$TIC_2(\sigma)$ is monotonically increasing with increasing $\sigma$, and since the Turing instability conditions are satisfied under the condition where the region is not growing, this condition is also satisfied if $\sigma > 0$.

Therefore, when the growth of the region starts, in order for the previously formed Turing patterns to be maintained, the bifurcation point of the growth rate is predicted in the linear theory as follows:

$$\sigma_c = \begin{cases} \sqrt[4]{F^2 K} - (F+K), & K - F \geq \sigma_D \\ \sigma_D, & K - F < \sigma_D \end{cases} . \tag{B12}$$

In addition, regarding the Bogdanov-Takens (BT) point, $\operatorname{tr} \hat{S}^* = 0$ and $\det \hat{S}^* = 0$ are required, leading to

$$\sigma_{BT} = \sigma_D = \sigma_T = \sqrt[4]{F^2 K} - (F+K) . \tag{B13}$$

Solving $D(\sigma_T) = 0$ yeilds

$$F = 4\sqrt{K^3} \text{ and } \sigma_{BT} = K - F. \tag{B14}$$

## C. Details for the Turing Space $(F, K, \sigma_c)$

Fig. 5 shows only the distribution on the $(F,K)$ plane and the maximum growth rate bifurcation point $\sigma_c$. The values of the growth rate bifurcation point $\sigma_c$ for each $(F,K)$ are plotted, and the gradient of the growth rate bifurcation points $\sigma_c$ depending on the $(F,K)$ plane is confirmed.

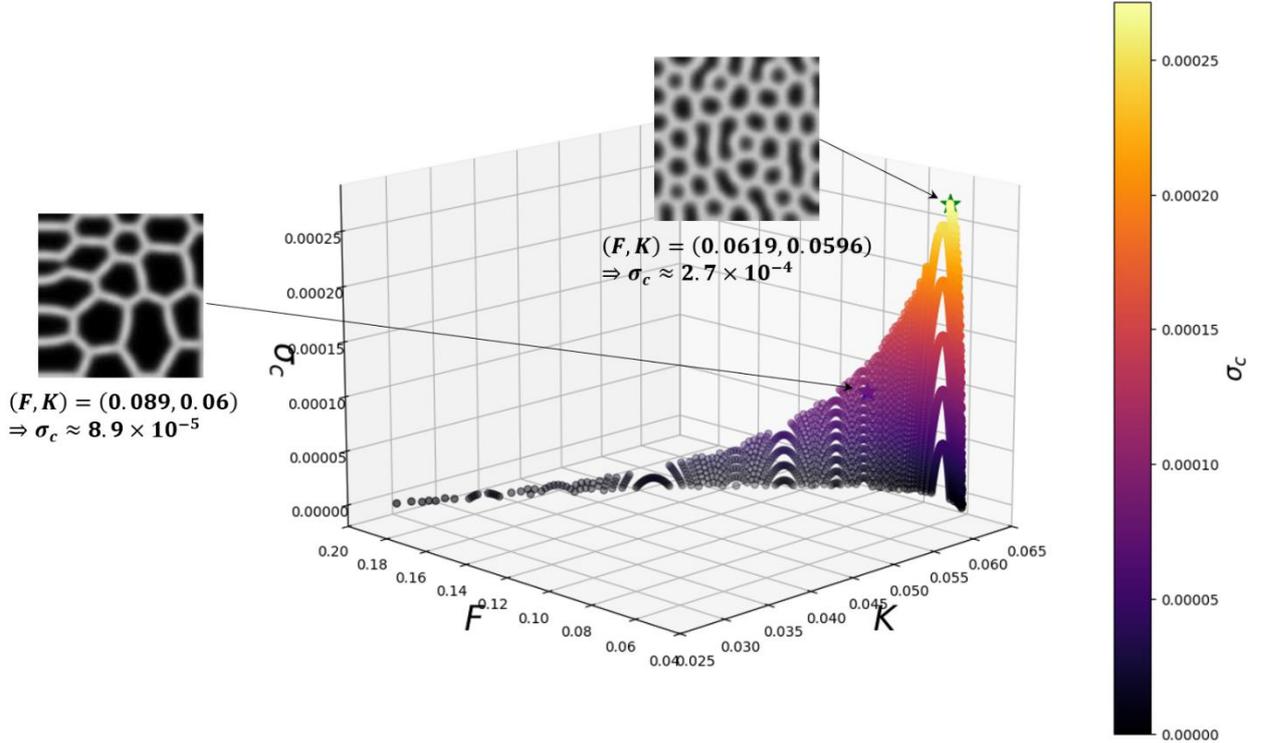

$(F, K) = (0.089, 0.06)$
$\Rightarrow \sigma_c \approx 8.9 \times 10^{-5}$

$(F, K) = (0.0619, 0.0596)$
$\Rightarrow \sigma_c \approx 2.7 \times 10^{-4}$

**Figure C:** In the figure, the growth rate bifurcation points $\sigma_c$ is plotted for pairs $(F,K)$ that satisfy the Turing instability condition, with $\phi_u$=0.2 and $\phi_v$=0.1 fixed in the reaction-diffusion system on a fixed domain. The left figure of Fig. 5 is rendered in 3D, and the gradient of the growth rate bifurcation points $\sigma_c$ for each $(F,K)$ is represented. The blue and green stars in the figure represent $\sigma_c$ in the equation (15) and $\sigma_{robust}$, respectively. (Figure in Color)

## D. Details for Figures 1, 2, 3, 4, 6, and 7

Using 3D graphs corresponding to Figures 1, 2, 3, 4, 6, and 7, we confirm the concentration gradient.

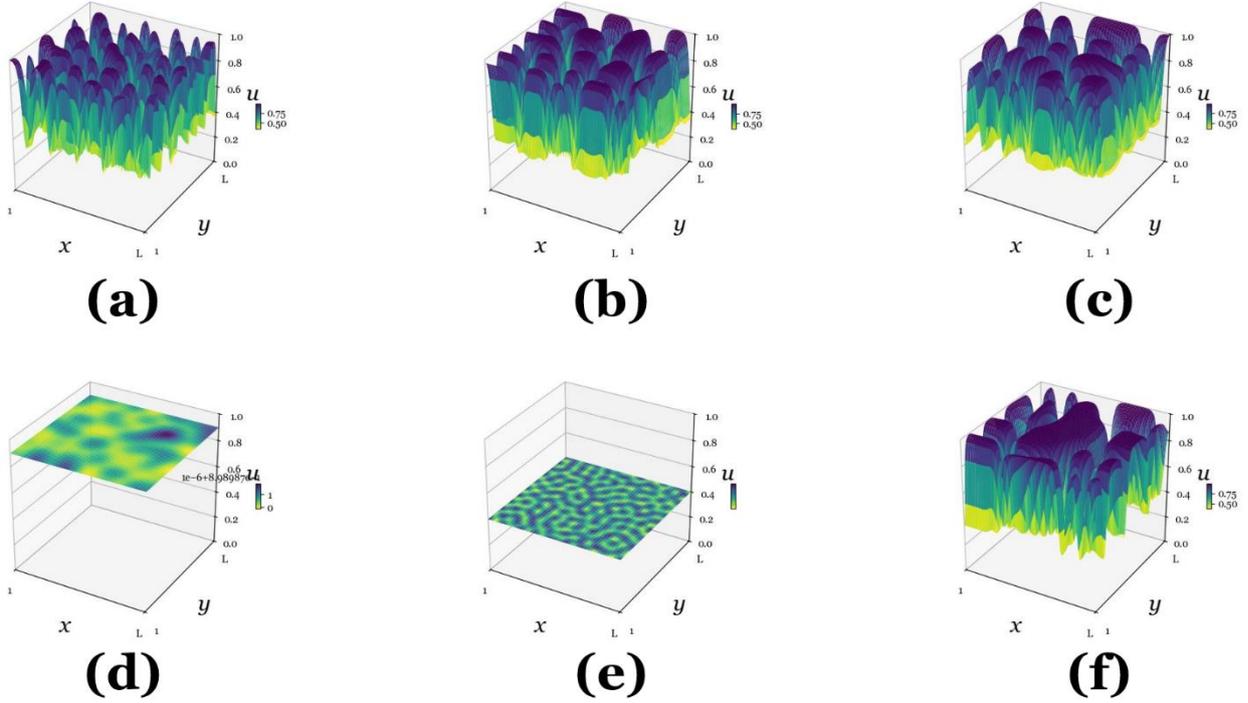

**Figure D**: The above (a), (b), (c), (d), (e), and (f) correspond to Figures 1, 2, 3, 4, 6, and 7, respectively. (Figures in Color)

### E.  Transition of the Reaction and Dilution Terms Values

We track values of the reaction and dilution terms in the system (5) at a certain fixed point, $(x, y) = (L/2, L/2)$, in the domain, in contrast to Fig. 8. As illustrated below, the further $\sigma$ is away from the growth rate bifurcation point $\sigma_c$, the faster the values of the reaction and dilution terms converge to zero, but in the vicinity of $\sigma_c$, the values fluctuate (in particular, transitions across signs are also observed in the cases: (**a**) and (**b**)), suggesting that the fixed point may stay in vicinity of the pattern formed.

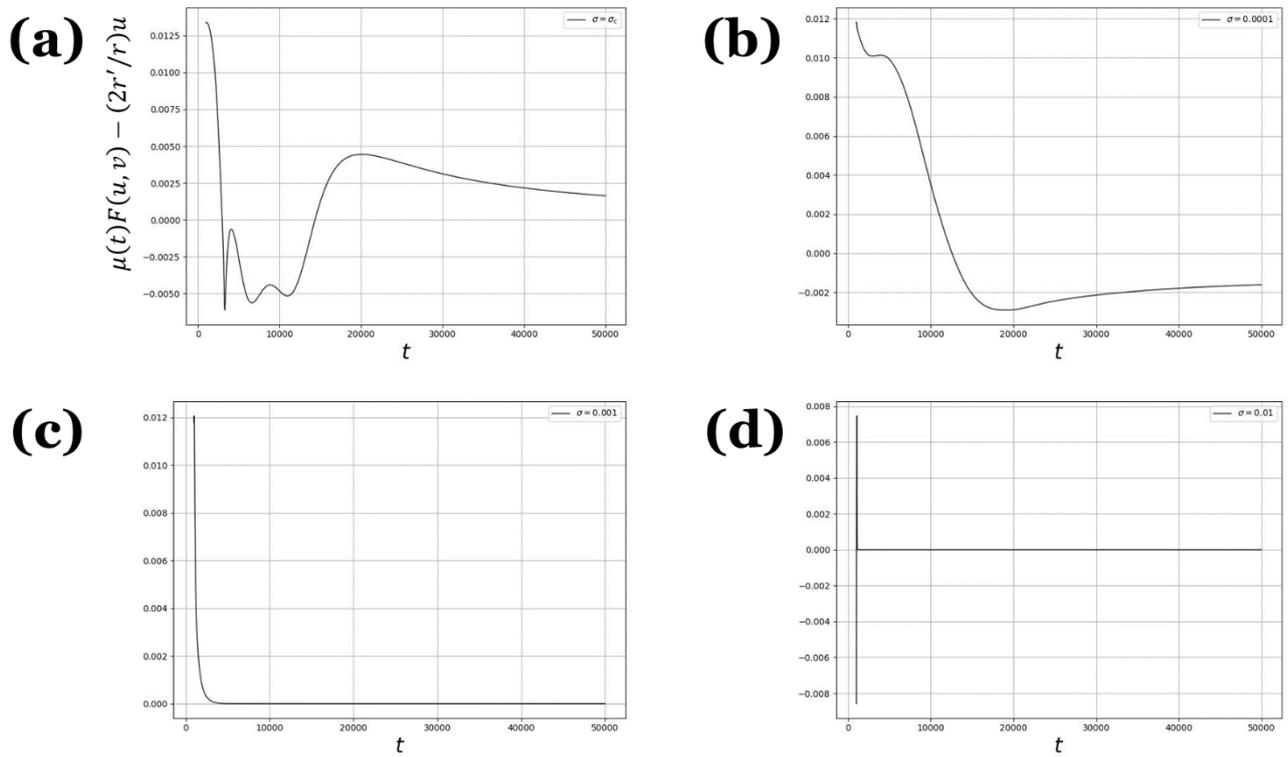

**Figure E**: Numerical analysis results of tracking $\mu(t)F(u,v) - (2r'/r)u$ in the system (5) at the fixed point $(x,y) = (L/2, L/2)$ until $t = 50{,}000$: (**a**) $\sigma = \sigma_c \approx 0.000089$, (**b**) $\sigma = 0.0001$, (**c**) $\sigma = 0.001$, (**d**) $\sigma = 0.01$.